\documentclass[10pt]{iopart}
\usepackage{graphicx}
\usepackage{xcolor}
\usepackage[export]{adjustbox}

\begin{document}

\title[Characterization of microscopic ferromagnetic defects in thin films using NV imaging]{Characterization of microscopic ferromagnetic defects in thin films using magnetic microscope based on Nitrogen-Vacancy centres}

\author{A. Berzins$^1$, J. Smits$^{1,2}$, A. Petruhins$^3$}

\address{$^1$Laser Centre, University of Latvia, Latvia}
\address{$^2$The University of New Mexico, Albuquerque, United States}
\address{$^3$Materials design group, Thinfilm physics division, Department of Physics,
Chemistry and Biology (IFM), Linkoping University, Sweden}

\ead{andris.berzins@lu.lv}
\vspace{10pt}
\begin{indented}
\item[]June 2020
\end{indented}

\begin{abstract}
In this work we present results acquired by applying magnetic field imaging technique based on Nitrogen-Vacancy centres in diamond crystal for characterization of magnetic thin films defects. We used the constructed wide-field magnetic microscope for measurements of two kinds of magnetic defects in thin films. One family of defects under study was a result of non-optimal thin film growth conditions. The magnetic field maps of several regions of the thin films created under very similar conditions to previously published research revealed microscopic impurity islands of ferromagnetic defects, that potentially could disturb the magnetic properties of the surface. The second part of the measurements was dedicated to defects created post deposition - mechanical defects introduced in ferromagnetic thin films. In both cases, the measurements identify the magnetic field amplitude and distribution of the magnetic defects. In addition, the magnetic field maps were correlated with the corresponding optical images. As this method has great potential for quality control of different stages of magnetic thin film manufacturing process and it can rival other widely used measurement techniques, we also propose solutions for the optimization of the device in the perspective of high throughput.

\end{abstract}

%
\vspace{2pc}

\submitto{\ MATERIALS CHEMISTRY AND PHYSICS}

%
\ioptwocol

\section{Introduction}

The Nitrogen-Vacancy (NV) centres in diamond crystal have potential for  for variety of applications, from magnetometry~\cite{wickenbrock_microwave-free_2016}, thermometry~\cite{clevenson_broadband_2015} and measurements of electric field~\cite{michl_robust_2019,dolde_electric-field_2011} to strain analysis~\cite{kehayias_imaging_2019}. An ensemble of NV centres that is fixed in concrete place in a diamond crystal can also be used to acquire 2D maps of the aforementioned physical properties.

Using NV centres for magnetic field imaging has a number of advantages: possibility to measure relatively wide area of sample simultaneously, while maintaining diffraction limited spatial resolution. A diamond crystal itself is a convenient base for variety of measurement conditions, as it can be brought into close proximity to the sample, as the diamond matrix is chemically and mechanically durable, as well as non-toxic. Furthermore, measurements can be made over a range of temperatures from cryogenic to several hundreds of degrees Celsius ~\cite{waxman_diamond_2014, wang_coherent_2020, plakhotnik_luminescence_2010}. The combination of these properties allows to investigate magnetic properties of thin films and magnetic structures on a microscopic scale~\cite{schirhagl_nitrogen-vacancy_2014,pham_magnetic_2011}, potentially allowing to monitor the manufacturing process of various structures as well as quality control of the final product.

The magnetic imaging could be useful for a number of thin film and microstructure applications~\cite{scheunert_review_2016} as it is based on standard microscopy techniques and therefore allows to combine optical images with magnetic field images. Examples for this are estimation of magnetic moments of microscopic magnetic particles~\cite{smits_estimating_2016}, magnetic domain topology~\cite{chesnel_morphological_2018}, ferromagnetic thin film growth~\cite{parveen_room-temperature_2017,liu_wafer-scale_2017}, visualization of magnetic leakage fields~\cite{shamonin_magneto-optical_2001} and orientation of liquid crystals~\cite{rupnik_field-controlled_2017}.

In this research we focused on imaging of ferromagnetic structures for several reasons: structural defects in ferromagnetic thin films tend to create magnetic defects with complicated patterns, ferromagnetic materials exhibit complicated properties even at low external magnetic fields and at room temperature, and ferromagnetic thin films have large variety of practical applications. For example, research of magnetic phases and their dependence of material thickness~\cite{ramirez_camacho_superparamagnetic_2020}, magnetic property research by material contents and processing~\cite{salameh_effects_2020}, wide range of perovskite structures~\cite{ramesh_multiferroics_2007,wakabayashi_ferromagnetism_2019, alam_room_2020}, magnetic shape memory thin films~\cite{golub_magnetism_2020, takhsha_ghahfarokhi_martensite-enabled_2020, pawar_magnetic_2020} as well as resistive random access memory~\cite{lee_electric_2020}, to name a few.

To demonstrate the capabilities of the magnetic field imaging technique we applied it to observe magnetic defects on a thin film produced under conditions very similar to previously published research. We found out that ferromagnetic signal comes from impurity defects condensed in microscopic islands as a result of non-optimal thin film growth process. Another group of measurements were done in ferromagnetic thin films which were exposed to mechanical interaction.

\section{Methods used}

\begin{figure}[ht!]
  \begin{center}
    \includegraphics[width=0.35\textwidth]{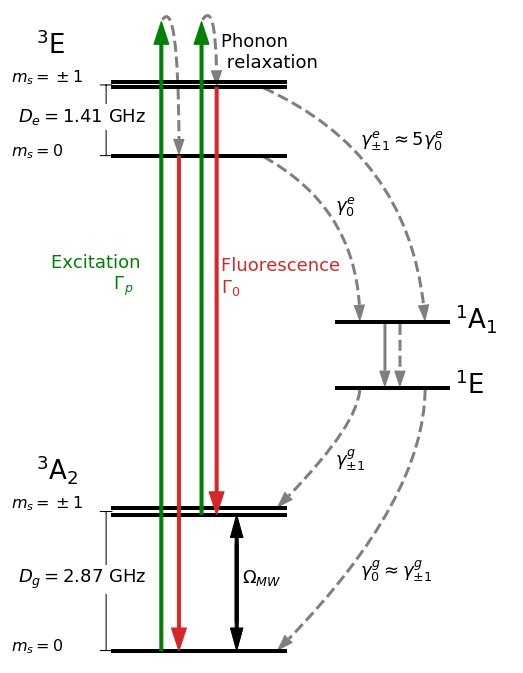}
  \end{center}
  \caption{Level scheme of an NV center in a diamond crystal, $m_S$ is the electron spin projection quantum number, $D_g$ and $D_e$ are the ground-state and excited-state zero-magnetic-field splittings, $\Omega_{MW}$ is the MW Rabi frequency, $\gamma_0^g$ and $\gamma_{\pm 1}^g$ are the relaxation rates from the singlet state $^1$E to the triplet ground state $^3$A$_2$, $\gamma_0^e$ and $\gamma_{\pm 1}^e$ are the relaxation rates from the triplet excited state $^3$E to the singlet state $^1$A$_1$ \cite{auzinsh_hyperfine_2019}.}
  \label{levels}
\end{figure}

To measure the properties of ferromagnetic thin films we used an optically detected magnetic resonance (ODMR) method. In FIG. \ref{levels} one can see the NV energy scheme which provides the basis for the measurement of ODMR signals. A frequency doubled Nd:YAG laser light (532 nm) is used to optically excite the NV centers. After light absorption with absorption rate $\Gamma_p$ and rapid relaxation in the phonon band the population of the excited state magnetic sub-levels can either decay back to the ground state with equal rates $\Gamma_0$ and radiate light in the red part of the spectrum or undergo non-radiative transitions to the singlet level $^{1}A_1$. These non-radiative transitions occur approximately five times more frequently from the excited state electron spin sublevels $m_S = \pm 1$ compared to $m_S = 0$. After the first step the singlet--singlet transition ${^{1}A}_1 \longrightarrow$ $^{1}E$ takes place with almost all the energy being transferred in a non-radiative way with a small fraction radiated in form of IR radiation. Finally, the population undergoes non-radiative transitions from $^{1}E$ to the ground triplet state with approximately equal transition probabilities to all three electron spin projection components of the $^3A_2$ level. Although the literature data is inconsistent and the relaxation rates vary in rather wide range~\cite{auzinsh_hyperfine_2019, jaskula_superresolution_2017, dumeige_magnetometry_2013}, in all cases the differences of the non-radiative transition rates in the excited triplet state $^3E$ leads to the situation that after several excitation--relaxation cycles the population of the NV centers in the ground triplet state will be transferred to the magnetic sublevel $m_S = 0$ and the electron spin angular momentum will be strongly polarized \cite{doherty_nitrogen-vacancy_2013}. In our measurements we used continuous laser excitation.

The optical polarisation can be used to measure magnetic field in the following way. The transitions from the ground state $m_S = \pm 1$ states due-to the larger probability of population relaxation to the non-radiative transitions results in lower luminescence signal. If the system starts with an optically polarized ground state (population in $m_S = 0$) and we add microwave frequency scan that at some point energetically connects the $m_S = 0$ and $m_S = -1$ or $m_S = +1$ levels, we will observe a drop in luminescence signal at the exact resonance frequency. This in order gives the distance between the ground state levels, and can be used for magnetic field measurements as the $m_S = \pm 1$ split in the external magnetic field. If the external magnetic field is aligned along one of the NV axes the $m_S = \pm 1$ states of the corresponding NV direction split linearly with growing magnetic field with $28.025$~MHz/mT ~\cite{doherty_theory_2012}.

In the case of magnetic field imaging, the luminescence signal must be collected and detected using an optical system, that maintains the two-dimensional information of the luminescence distribution of the NV layer. After the data processing (discussed in next chapter) this gives the magnetic field distribution over the field-of-view.

\section{Data processing}

\begin{figure}
      \includegraphics[width=0.5\textwidth]{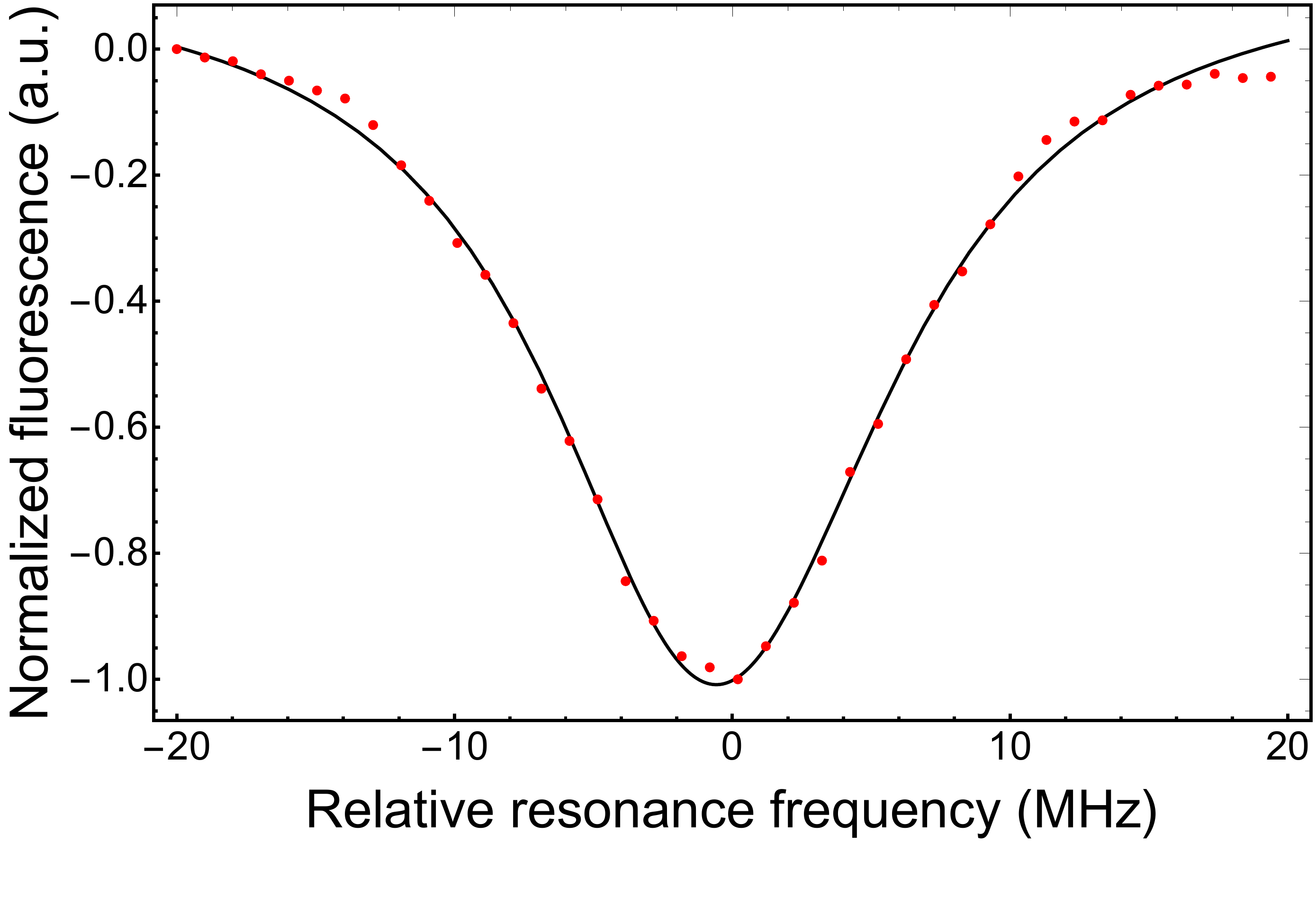}
    \caption{An example of data for one camera pixel: the experimentally measured data points (red circles), fitted with a combination of three Lorentzian profiles (black line).}
  \label{data}
\end{figure}

The acquired data (full ODMR shape for each pixel) was processed in the following way. The ODMR signal was fitted with a combination of three Lorentzian profiles (corresponding to each of the three hyperfine components) for each pixel, where each experimental point represents a narrow MW field value. This is done to account for the asymmetry in the line-profile (see Figure~\ref{data}) due to polarization of the nitrogen nucleus forming the NV center~\cite{doherty_nitrogen-vacancy_2013, neumann_excited-state_2009}. The obtained fit values for the transition frequencies (minimum value of the fit curve) were plotted as a 2D heat-map revealing the spatial distribution of the magnetic field.

Determining the minimum value from the fitted curve lessens the imperfections in data quality as noise and relatively sparse data points at the tip of the resonance shape. In this case seemingly the nuclear spin polarisation plays small role in the shape of the ODMR profile, but it has to be noted that the shape and FWHM of the resonance is this wide due-to the specifics of the thin film samples - the local magnetic field changes rapidly and even one pixel of the camera detects NV centres with different resonance frequencies. For comparison the same diamond crystal and set-up with small modifications were used to measure other type of samples~\cite{berzins_investigation_2020}, and gave notably narrower FWHM and much more pronounced signature of the nuclear spin polarisation. The MW power and magnetic bias field used in the measurements were roughly the same, but the laser power in this research was two times smaller.

\section{Experimental Device}

\begin{figure}
      \includegraphics[width=0.5\textwidth]{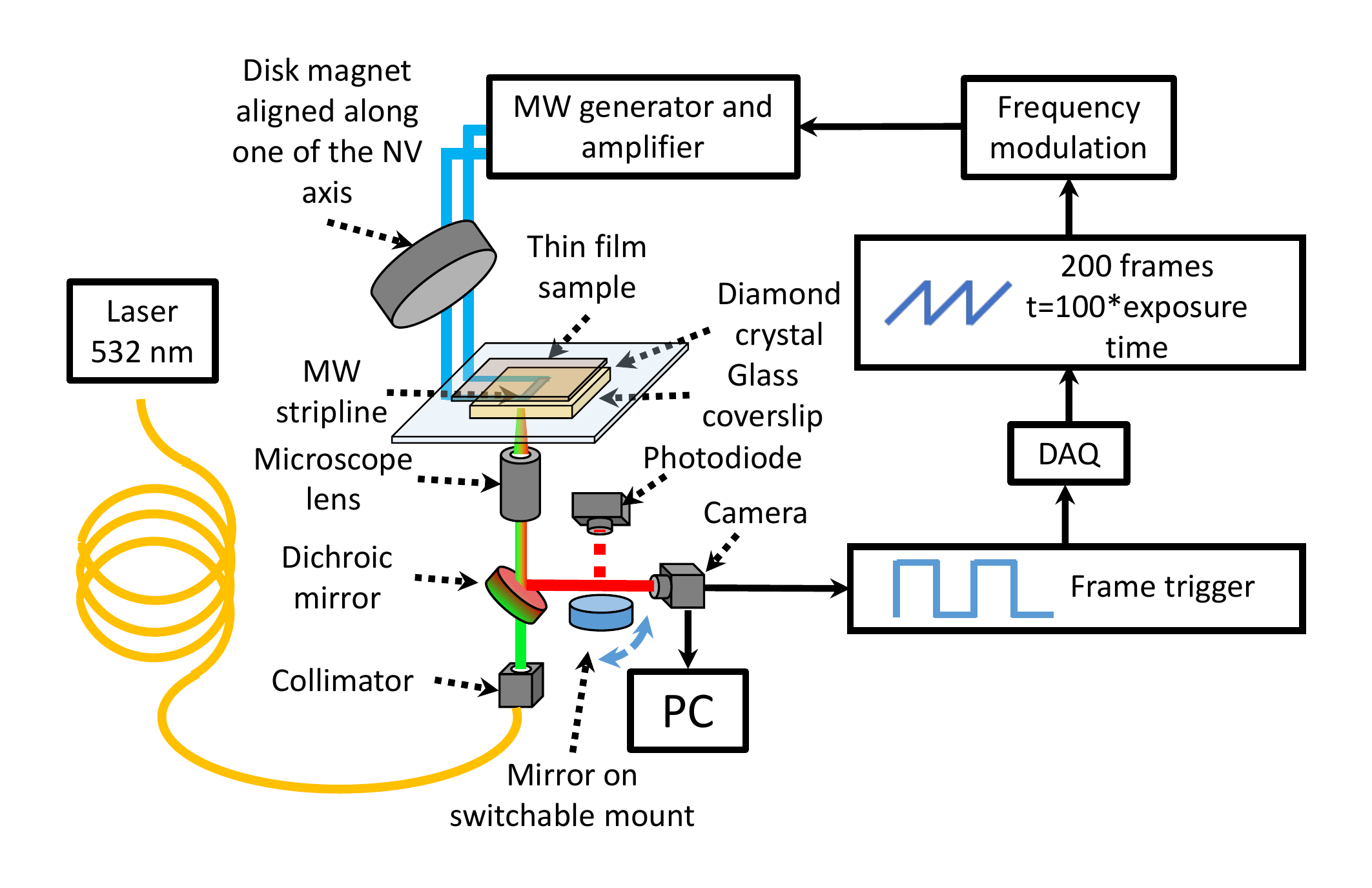}
    \caption{Experimental setup. The laser light was coupled into an optical fiber that led it to the experimental system, in which a dichroic mirror (Thorlabs DMLP567R) reflects the green light, which, in turn, was directed to the sample via optical system that insured smooth distribution of exciting radiation across the field of view. The luminescence from the sample was collected with a 40x infinity-corrected objective with an NA of 0.65, and, after passing through the dichroic mirror a nd a long-pass filter (Thorlabs FEL0600), it was focused onto the sCMOS matrix of the camera (Andor NEO 5.5) or a photodiode (Thorlabs PDA36A-EC)}
  \label{setup}
\end{figure}

The experimental scheme for magnetic field imaging device is depicted in Figure~\ref{setup}. We used a Coherent Verdi V-18 laser to excite the NV centres. Laser power at the sample was around 40 mW. We used a dynamic transmissive speckle reducer (Optotune LSR-3005) to suppress interference artifacts (very pronounced in cases with well reflecting samples)  originating from the thin air gap between the diamond and the sample. A lens system was used to optimise the lighting over the field of view. In our case the field-of-view was a 110x110 $\mu m^2$ square. We used an epifluorescent set-up where the excitation and luminescence detection are done through the same optical path and the red luminescence was separated by dichroic mirror (see Fig. \ref{setup}). After that the luminescence was detected by either a sCMOS sensor of Andor Neo 5.5 camera, or a photodiode (Thorlabs PDA36A-EC). The MW field was delivered to the NV centres by using a copper stripline on a glass slide. The shape of the stripline was a straight wire (in all measurements the distance of point of interest and the MW stripline is 200 micrometers or less). Measurements of two dimensional ODMR maps were done by sweeping the MW frequency while simultaneously acquiring a series of camera frames. The microwave frequency modulation was controlled by an analog voltage generated by a data acquisition card (National Instruments USB-6001). The sweep was triggered by a pulse generated by the camera itself and a fixed burst of frames were acquired in sync with the modulation waveform (a sawtooth wave). A 40 frames were acquired for each sweep and 5 sweeps were done between each readout. The 5 separate sweeps were then averaged together. This was done to minimize down-time from shuttling data in memory. In all measurements the external magnetic field was aligned along one of the NV axes (100 orientation), meaning that it is oriented at around 35$^{\circ}$ to the diamond surface. The thin film samples under study were sputtered on a MgO base with (111) surface orientation. The samples were put directly on the diamond crystal and the gravity was the only force that pushed the samples down.

The diamond sample we used (bought from Element 6) was an electronics grade CVD diamond with a (100) surface polish. The physical dimensions of the crystal are 3 mm $\times$ 3 mm $\times$ 0.1 mm. The crystal was irradiated with nitrogen ions at three separate energies: 10 keV, 35 keV and 60 keV and then annealed at high temperature so that the vacancies created by the irradiation could migrate and form NV centres. The SRIM simulation for this diamond sample is already published before and can be found in~\cite{smits_estimating_2016}, and it gives a distribution of vacancies in depth of around 100 nm. Due to some uncertainty in the final distribution of NV centres after annealing we estimate that the NV layer depth is between 100-200 $nm$ thick and it is located just below one of the surfaces of the diamond crystal.

\section{Results and discussion}

To demonstrate defect detection technique using NV centres and ODMR approach we used two types of ferromagnetic thin films, one being a thin film with ferromagnetic impurities resulting from non-optimal growth conditions and the second being a ferromagnetic films with \textit{ex situ} introduced mechanical defects. All of the measurements were performed at room temperature.

In the first case, nanolaminated (atomically layered) thin films of (Cr$_{0.5}$Mn$_{0.5}$)$_2$GaC were chosen, whose properties and deposition conditions are described elsewhere~\cite{petruhins_synthesis_2015}. This thin film was of specific interest, as similar thin films were measured before using vibrating sample magnetometry, ferromagnetic resonance and SQUID magnetometry ~\cite{petruhins_synthesis_2015,salikhov_magnetic_2015}. However, even grown under optimal or near-optimal conditions, produced films can contain a small amount of surface impurities, which in turn can produce a ferromagnetic signal in addition to the signal arising from main phase present in the thin film. The magnetic field imaging technique based on NV centres allows to pinpoint the existence and origin of these signals, as well as to identify the magnetic properties of these structures.

\begin{figure*} 
\begin{center}
\includegraphics[width=0.43\textwidth,valign=t]{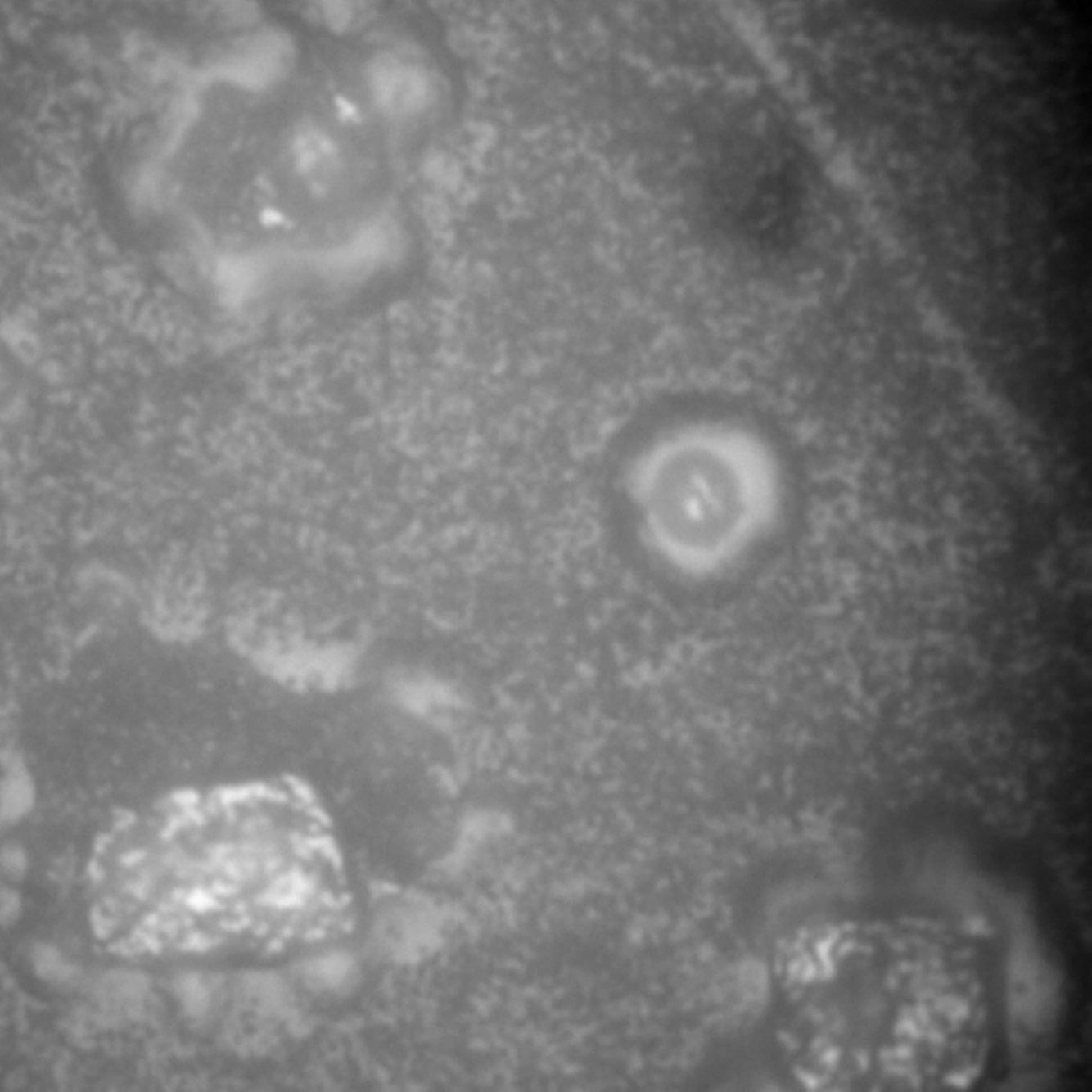}
\hspace{0.1cm}
\includegraphics[width=0.45\textwidth, trim={0cm 1cm 0cm 0cm},clip,valign=t]{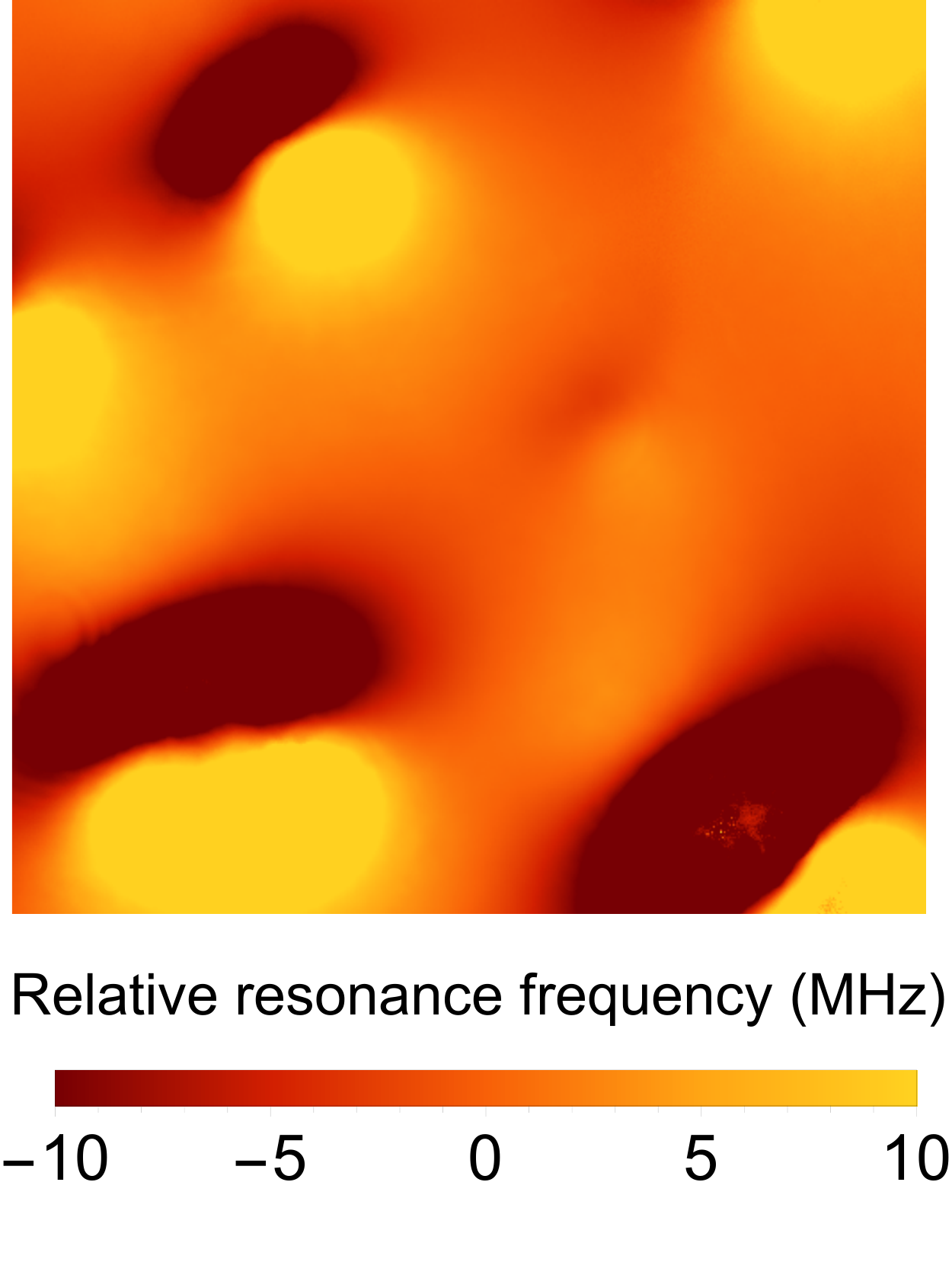} 
\end{center}
\caption{Measurements of (Cr$_{0.5}$Mn$_{0.5}$)$_2$GaC thin film surface defects created due-to non-optimal thin film growth process. \textbf{Left panel}: optical image taken using the same optical system but with white light illumination. The lighter material covering almost whole surface of the field of view is Gallium (nonmagnetic). The islands forming the local magnetic field defects are attributed to impurities consisting of Mn$_5$Ga$_8$, Cr$_5$Ga$_8$ or solid solution of the constituents, i.e. (Cr$_{1-x}$Mn$_x$)$_5$Ga$_8$. \textbf{Right panel}: The ODMR map representing the magnetic field changes over the field of view. The small imperfection at the bottom-right corner of the ODMR map arises from the resonance frequency exiting the scanning region}
\label{surface_1}
\end{figure*}

\begin{figure*} 
\begin{center}
\includegraphics[width=0.415\textwidth, trim={0cm 0cm 0cm 0cm},clip,valign=t]{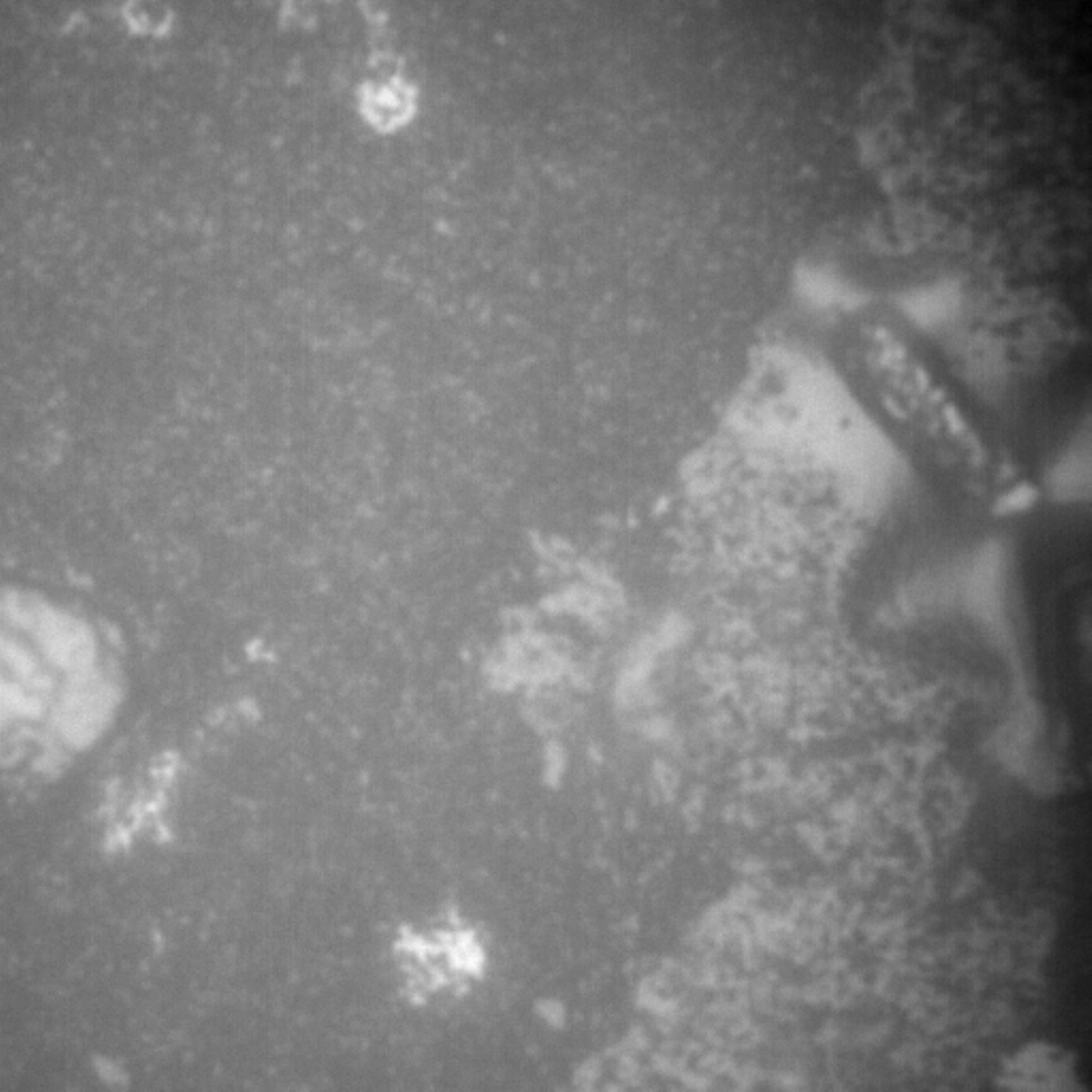}
\hspace{0.1cm}
\includegraphics[width=0.45\textwidth, trim={0cm 1cm 0cm 0.5cm},clip,valign=t]{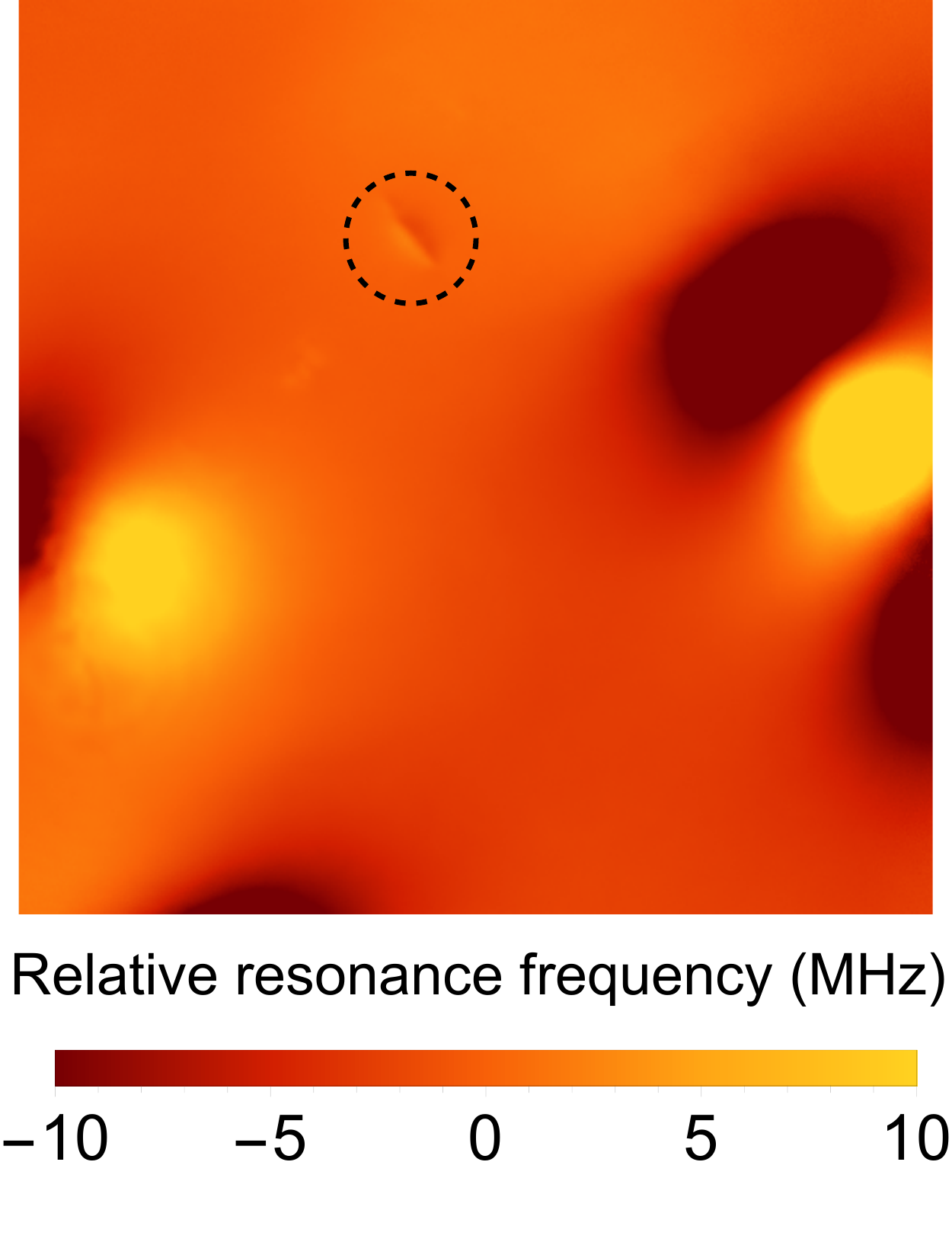} 
\end{center}
\caption{Measurements of (Cr$_{0.5}$Mn$_{0.5}$)$_2$GaC thin film surface defects created due-to non-optimal thin film growth process. \textbf{Left panel}: optical image taken using the same optical system but with white light illumination. The lighter material prominently covering right half of the field of view is Gallium (nonmagnetic). The islands forming the local magnetic field defects are attributed to impurities consisting of Mn$_5$Ga$_8$, Cr$_5$Ga$_8$ or solid solution of the constituents, i.e. (Cr$_{1-x}$Mn$_x$)$_5$Ga$_8$. \textbf{Right panel}: The ODMR map representing the magnetic field changes over the field of view. The defect in the dashed circle is related to inner strain of the diamond crystal, and not to the properties of the thin film.}
\label{surface_2}
\end{figure*}

The first set of measurements (Figures~\ref{surface_1} and~\ref{surface_2}) are depicting the surfaces of a thin film (field of view 110x110 $\mu m^2$): the left-side panels present optical images acquired by illuminating the sample with a white light source, and the right-side panels present the ODMR maps (magnetic field images), where the color scale represent a relative shift of the resonance frequency. In both cases there are a very distinct structures with varying shapes that create difference in relative magnetic field shifts approaching 1 $mT$. These shapes are most likely attributed to impurities consisting of Mn$_5$Ga$_8$, Cr$_5$Ga$_8$ or solid solution of the constituents, i.e. (Cr$_1-x$Mn$_x$)$_5$Ga$_8$, forming islands on the main thin film, as they were clearly seen in X-ray diffraction in other films during early stages of film optimization (not published). Although these shapes my resemble grains on the surface, these structures are relatively flat as they are in the optical focus at the same time as the surface of the main body of the thin film is. The non-magnetic structure, that can be seen in both optical images is pure gallium (lighter shade covering relatively large surface areas). Interestingly, the islands (prominently seen in optical image of Fig.~\ref{surface_1}) seem to strip away some part of the gallium from the surface to form the island. We deduce that the element on the surface of the thin film is gallium, as it has been demonstrated in similar nanolaminated thin films ~\cite{eklund_epitaxial_2011,ingason_magnetic_2013} that the \textit{A} element of the \textit{M}$_{n+1}$\textit{A}\textit{X}$_n$ phases~\cite{barsoum_mn1axn_2000} tend to segregate to the surface of the film.

It has to be noted that at the edges of the magnetic images one can see some magnetic structures, that can not be identified in the optical images - these patterns occur due to defects outside field-of-view, but as the created magnetic field extends beyond the physical dimensions of magnetic structures, one can see them in magnetic images.

It should be noted that the analysis performed in publications ~\cite{petruhins_synthesis_2015} and ~\cite{salikhov_magnetic_2015} was done on thin films grown under optimal or near optimal conditions, whereas the thin films analysed in present work have been synthesized under very similar, but slightly differing conditions, thus leading to surface impurities, which were present on some parts of the film surface. Another aspect that must be stressed here is the magnetic properties of interest. While the two papers~\cite{petruhins_synthesis_2015} and ~\cite{salikhov_magnetic_2015} concentrated on the magnetic properties of the thin film itself (prominently revealed at cryogenic temperatures), we focused on the surface properties that might be relevant for some applications.

In the upper central part of the ODMR map of Fig.~\ref{surface_2}, denoted with a dashed circle, there is a defect that is not related to the magnetic properties of the thin film, but rather to a local defect within the diamond crystal lattice, this is a well known phenomenon~\cite{kehayias_imaging_2019, knauer_-situ_2020,trusheim_wide-field_2016}.

\begin{figure*} 
\begin{center}
\includegraphics[width=0.415\textwidth,valign=t]{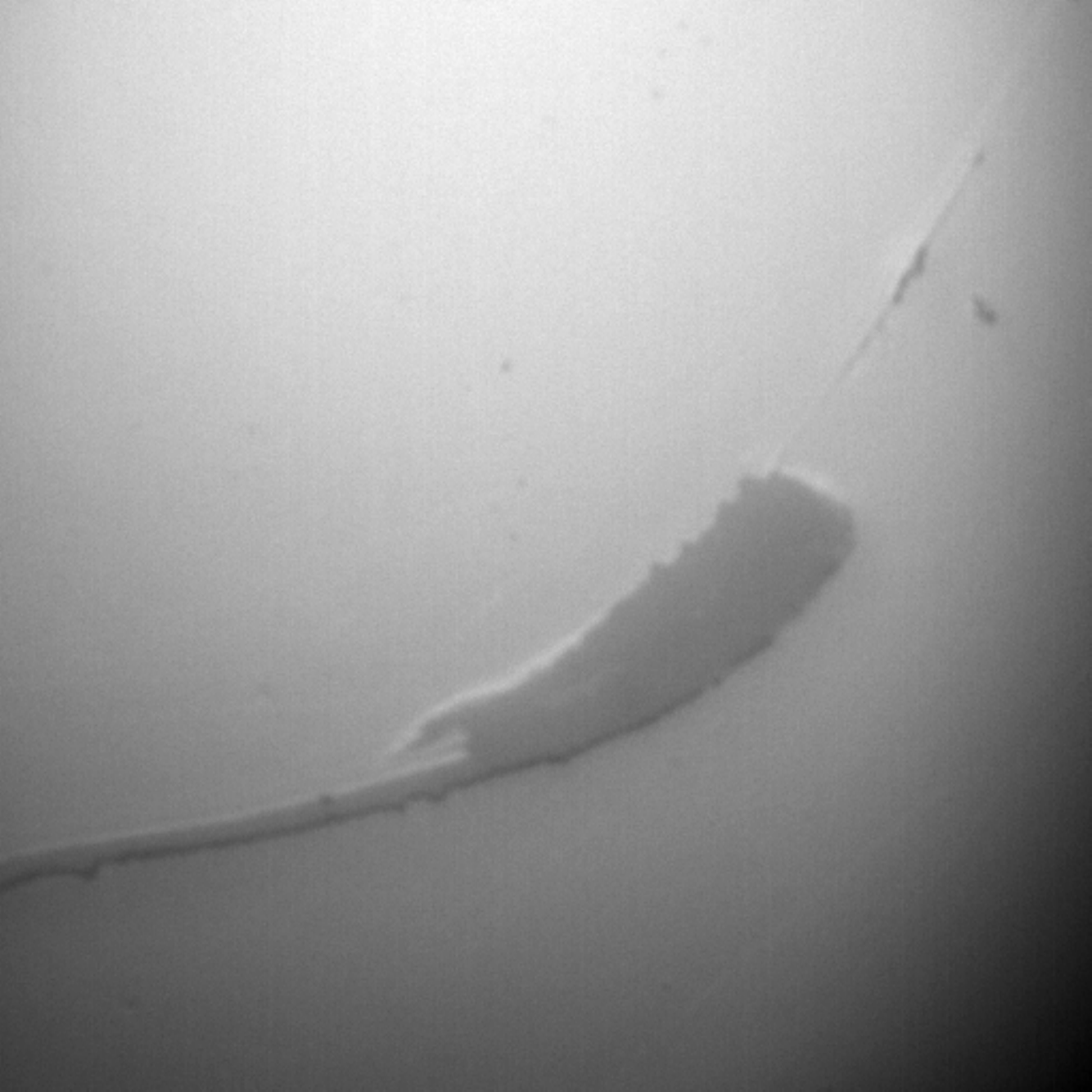}
\hspace{0.1cm}
\includegraphics[width=0.45\textwidth, trim={0cm 0.2cm 0cm 0.5cm},clip,valign=t]{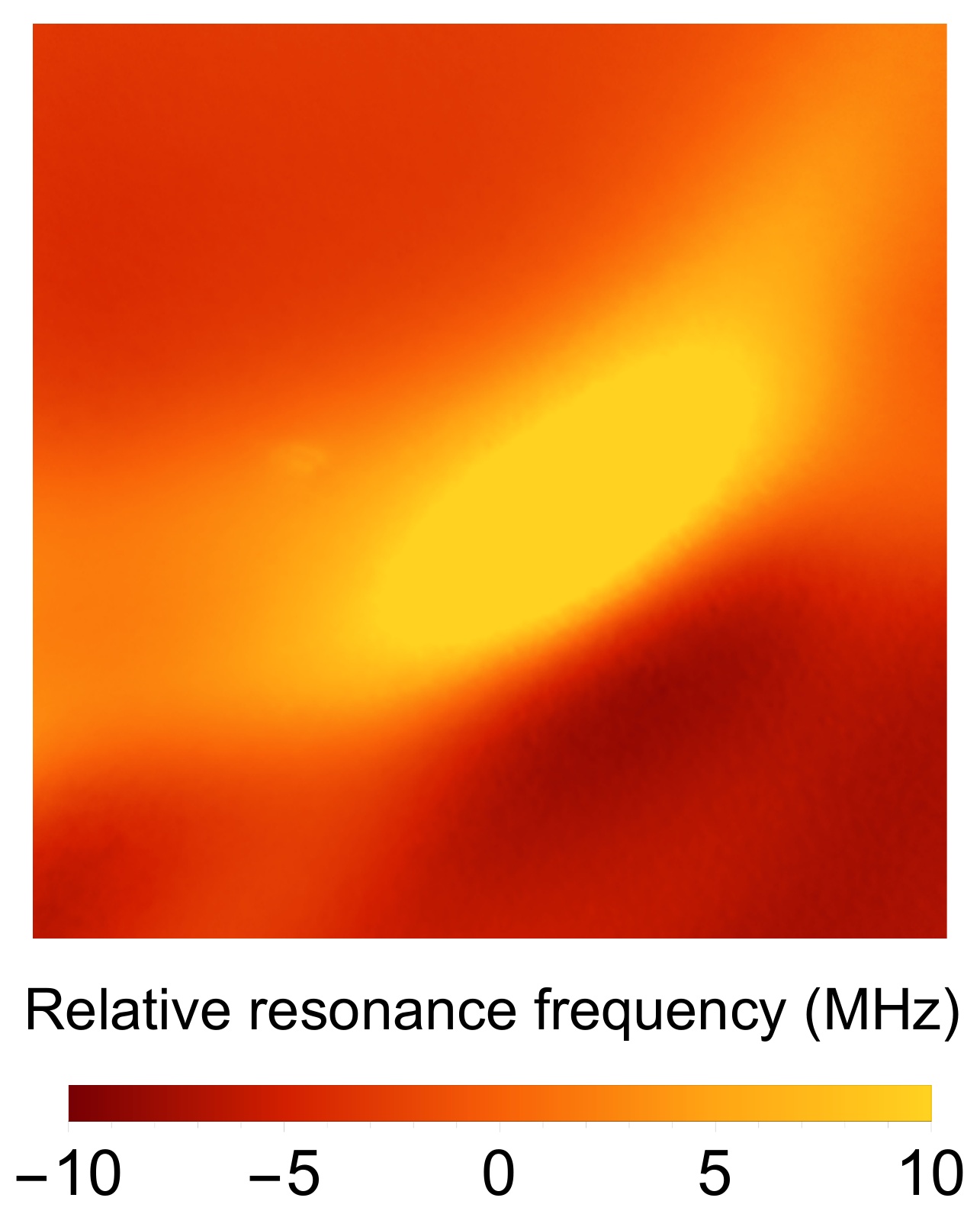} 
\end{center}
\caption{Measurements of 500 nm thick Fe thin film with a point like surface defect introduced by a needle punch. \textbf{Left panel}: optical image taken using the same optical system but with white light illumination. \textbf{Right panel}: The ODMR map representing the magnetic field changes over the field of view.}
\label{dot}
\end{figure*}

\begin{figure*} 
\begin{center}
\includegraphics[width=0.415\textwidth,valign=t]{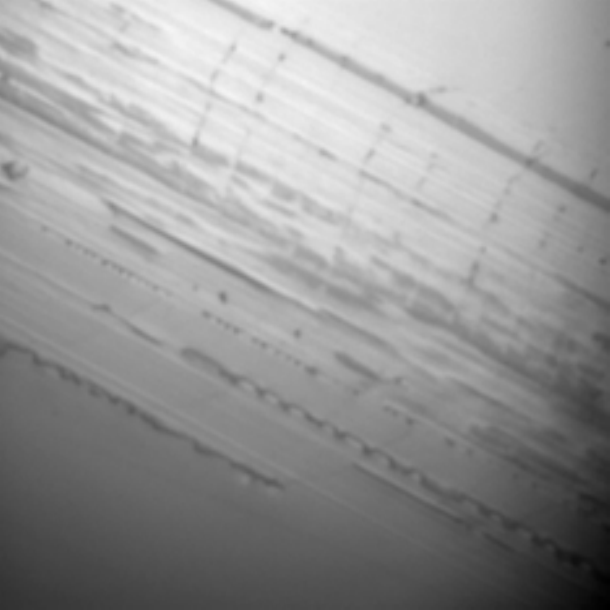}
\hspace{0.1cm}
\includegraphics[width=0.45\textwidth, trim={0cm 1cm 0cm 0.5cm},clip,valign=t]{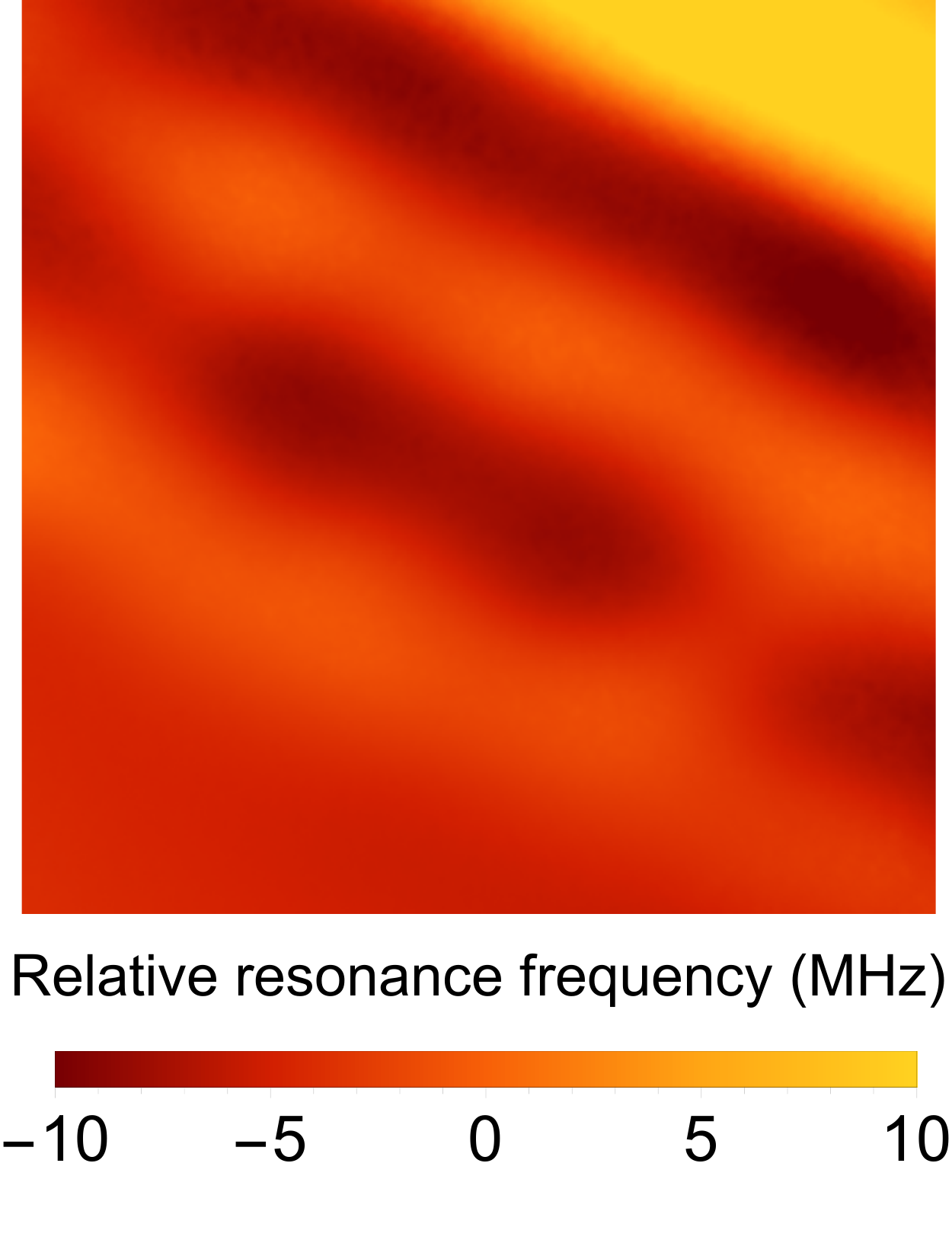} 
\end{center}
\caption{Measurements of 500 nm thick Fe thin film with line defects on the surface introduced by an assembly knife. \textbf{Left panel}: optical image taken using the same optical system but with white light illumination. \textbf{Right panel}: The ODMR map representing the magnetic field changes over the field of view.}
\label{stripe}
\end{figure*}

The second type of magnetic films that were analysed are samples with \textit{ex situ} mechanical defects on the 500 nm thick iron thin film, whose magnetic images are shown in Fig.~\ref{dot} (needle punch) and Fig.~\ref{stripe} (defect introduced by assembly knife, consisting of multiple lines). In both cases the magnetic signature of the thin film defect is clearly visible in contrast to the spots where the thin film is intact. As expected ferromagnetic thin film creates strong variations of local magnetic field (variations reaching 1 $mT$ difference) and spatially rapidly changing magnetic pattern at the spot where the integrity of the thin film is broken. In the case of surface defects introduced by an assembly knife (Fig.~\ref{stripe}) it is visible that the defect structure is relatively rich with scratches differing in width and depth that are not clearly distinguishable in the magnetic image. As in the previous case with the ferromagnetic impurity islands, the magnetic field creates a significantly larger magnetic pattern around the defect than its physical dimensions. 
This can be a useful effect if the magnetic field imaging technique is used for a quality control of thin film parameters (for example magnetic shape memory thin films  and resistive random access memory, mentioned in the Introduction), as it helps to magnify the the defect size, thus even a defect with dimensions well below the optical resolution could be clearly seen.

With examples that clearly demonstrate the capabilities of the magnetic field imaging technique let us discuss the optimisation perspectives of similar devices. To use magnetic field imaging for quality control of different steps of magnetic thin film fabrication, there can be various aspects that could be easily modified to upgrade performance of the device. The first thing is the measurement time - to acquire smooth magnetic field images, as shown in figures~\ref{surface_1}, ~\ref{surface_2}, ~\ref{dot} and~\ref{stripe} the measurement time was around 20 minutes per one ODMR map. To potentially identify the shift from the expected value, the measurement time could be reduced by many orders of magnitude. In our measurements the MW frequency was scanned across 40 MHz range (yielding 40 data points) to obtain full information about the ODMR shape. This parameter can be reduced to 3 concrete MW frequencies if the direction of the frequency shift is important, or to a two MW frequency mode if only deviation from concrete magnetic field value is monitored (one frequency to monitor signal level, other to monitor changes in contrast). Another factor determining the throughput of similar devices is the contrast of the ODMR signals. In this case we used very modest laser power, as our set-up was built to correlate the optical image with the magnetic image, thus the laser light reached the magnetic sample itself. If the surface of the diamond crystal would be covered in a reflective thin film, or the laser excitation was provided in the direction parallel to the surface of the magnetic thin films, one could use much larger laser power, insuring better contrast of ODMR signals. This would result in smaller number of required averages, that in order could easily reduce the measurement time by two orders of magnitude.

One more aspect that needs to be considered is the properties of the possible defects to be studied. The advantage of magnetic detection of defects in ferromagnetic thin films lies within the fact, that in contrary to defects studied in this work, structural imperfections can occur within the volume of a thin film, with no visible surface deformation. In such case the magnetic structure will still be detectable, while the visual surface monitoring methods will fail to detect the defect.

\section{Conclusions}

We have demonstrated the experimental results that clearly identify the different magnetic properties of structures on the surface of thin films caused by two types of surface defects: surface impurities formed during the thin film growth and defects created by mechanical interaction with the thin film. The studied samples shows the strengths and advantages of wide field magnetic microscopy with high spatial resolution in detection of local magnetic defects. This was clearly demonstrated by pinpointing and describing the presence of ferromagnetic impurities, attributing the ferromagnetic behavior to microscopic islands on the surface of the thin film. The mechanical defects on a solid iron thin film were also described successfully, showing the complex magnetic structure created by damaged integrity of the thin films. We have also discussed the first steps of optimization of similar systems towards higher throughput. In comparison to widely used vibrating sample magnetometry the magnetic field imaging method based on NV centres delivers sub-micrometer scale spatial resolution, opening up new possibilities in thin film research and quality control. In comparison to scanning tip magnetometry performed by superconducting quantum interference devices (SQUID) or single NV probes~\cite{appel_nanomagnetism_2019} the method presented here delivers a wide-field magnetic microscopy, that cannot compete in sensitivity, but is superior in measurement speed and potential throughput.

\section{Acknowledgements}

A. Berzins acknowledges support from PostDoc Latvia project 1.1.1.2/VIAA/1/16/024 "Two-way research of thin-films and NV centres in diamond crystal" and LLC "MikroTik" donation, administered by the UoL foundation, for opportunity to significantly improve experimental set-up.

\clearpage

\bibliography{references}
\bibliographystyle{unsrt}

\end{document}